\documentclass[runningheads]{llncs}
\usepackage[utf8]{inputenc}
\usepackage{amsmath,amssymb,bm,cancel,xcolor,bussproofs,multicol,hyperref,multirow,mathtools,tikz,array,mdframed}
\usetikzlibrary{positioning,arrows.meta}
\usepackage[only,llbracket,rrbracket,bbslash]{stmaryrd}
\usepackage{thm-restate}
\usepackage{xspace,bbding}
\usepackage{stackengine}
\usepackage{pifont}

\usepackage{pgfplots}
\pgfplotsset{compat=1.18}

\usepackage{todonotes}
\usepackage{wrapfig}
\usepackage[shortlabels]{enumitem}
\usepackage{cleveref}
\usepackage{bbding}

\usepackage{hyperref}

\makeatletter
\newcommand{\clab}[2]{%
   \protected@write \@auxout {}{\string \newlabel {#1}{{#2}{\thepage}{#2}{#1}{}} }%
   \hypertarget{#1}{#2}
}
\makeatother

\makeatletter
\newcommand\smaller{\@setfontsize\smaller\@viiipt\@ixpt}
\makeatother

\makeatletter
\newcommand{\leqnomode}{\tagsleft@true\let\veqno\@@leqno}
\makeatother

\spnewtheorem{Definition}{Definition}{\bfseries}{\normalfont}
\Crefname{Definition}{\protect Definition}{\protect Definitions}

\makeatletter
\spn@wtheorem{Lemma}{Lemma}{\bfseries}{\normalfont}
\makeatother

\Crefname{Lemma}{\protect Lemma}{\protect Lemmas}

\renewcommand{\paragraph}[1]{\par\medskip\noindent\textbf{#1}}

\newcommand{\eqs}{\approx}
\newcommand{\neqs}{\not\approx}
\newcommand{\deqs}{\mathrel{\dot\approx}}
\newcommand*{\defeq}{\stackrel{\scriptscriptstyle\textsf{def}}{=}}

\newcommand{\con}[1]{\setminus #1}
\newcommand\infers{\to}
\newcommand\rf{\ensuremath{R}\xspace}
\newcommand\Calc{PaRC}
\newcommand\review[1]{#1}
\newcommand\cameraready[1]{}
\newcommand\fullversion[1]{#1}

\newcommand{\mgu}{{\sf mgu}}
\def\orcidID#1{\href{http://orcid.org/#1}{\raisebox{-1.25pt}{\includegraphics{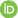}}}}
\newcommand{\CommentedOut}[1]{}

\newcommand{\CC}{\ensuremath{\mathcal{C}}}
\newcommand{\CD}{\ensuremath{\mathcal{D}}}
\newcommand{\satS}{\mathbb{S}} 
\newcommand{\gnr}[1]{#1^\circledast} 
\newcommand{\gnrS}{\gnr{S}}    
\newcommand{\setof}[1]{\{#1\}} 
\newcommand{\emptyclause}{\Box}
\newcommand{\QEDsymbol}{\text{\ding{111}}}
\newcommand{\QED}{\hspace*{\stretch{1}}\QEDsymbol}
\newcommand{\dotin}{\mathrel{\hat{\in}}} 

\newenvironment{Case}[2]%
    {\noindent\textit{\textbf{Case #1.} #2}\par\medskip}%
    {\medskip}
\newenvironment{Subcase}[2]%
    {\noindent\textit{\textbf{Subcase #1.} #2}\par\medskip}%
    {\medskip}

\newcommand\solver[1]{\textsc{#1}\xspace}
\newcommand\vampire{\solver{Vampire}}

\newcommand{\Sup}{{\sf Sup}}
\newcommand{\EqRes}{{\sf EqRes}}
\newcommand{\EqFac}{{\sf EqFac}}

\begin{document}

\title{Partial Redundancy in Saturation}
\author{M\'arton Hajdu\inst{1}\textsuperscript{(\Envelope)}\orcidID{0000-0002-8273-2613}
\and
Laura Kov\'acs\inst{1}\orcidID{0000-0002-8299-2714}
\and
Andrei Voronkov\inst{2,3}
}
\institute{TU Wien, Vienna, Austria\\
\email{marton.hajdu@tuwien.ac.at}
\and University of Manchester, Manchester, UK \and EasyChair, Manchester, UK}
\date{}

\authorrunning{M\'arton Hajdu, Laura Kov\'acs and Andrei Voronkov}

\maketitle
\begin{abstract}
Redundancy elimination is one of the crucial ingredients of efficient saturation-based proof search. We improve redundancy elimination by introducing a new notion of redundancy, based on \emph{partial clauses} and \emph{redundancy formulas}, which is more powerful than the standard notion: there are both clauses and inferences that are redundant when we use our notions and not redundant when we use standard notions.
In a way, our notion blurs the distinction between redundancy at the level of inferences and redundancy at the level of clauses. 
We present a superposition calculus \Calc\ on partial clauses. Our calculus is refutationally complete and is strong enough to capture some standard restrictions of the superposition calculus. 
We discuss the implementation of the calculus in the theorem prover \vampire. Our experiments show the power of the new approach: we were able to solve 24 TPTP problems not previously solved by any prover, including previous versions of \vampire.
\end{abstract}

\section{Introduction}
Modern theorem provers for first-order logic with equality use variants of the \emph{superposition calculus}~\cite{ParamodulationAndTheoremProving,NieuwenhuisRubio:HandbookAR:paramodulation:2001,EquationalReasoningInSaturation}. 
A key ingredient in superposition reasoning is the use of \emph{simplification orders} on terms and clauses. 
Efficiency is achieved by using a powerful concept of \emph{redundancy}. 
For example, it is known that a clause that is a logical consequence of smaller clauses in the search space can be removed from the search space. While checking redundancy is undecidable, there are special cases, such as demodulation, which are easy to check and are widely used. 
The efficiency of superposition is evidenced by the practical performance of saturation-based provers, including  E~\cite{E19}, \vampire{}~\cite{CAV13}, and iProver~\cite{iProver}.
\newpage
\review{
Standard completeness arguments of superposition calculi with redundancy are done by studying the behavior of the set of ground instances of clauses in the search space~\cite{NieuwenhuisRubio:HandbookAR:paramodulation:2001}. 
When all ground instances are redundant, the clause becomes redundant.
In this paper, we propose a powerful generalization of standard redundancy, called \emph{partial redundancy}. 
In a nutshell, the idea is the following. 
Each superposition inference produces a clause that makes some set of ground instances in the search space redundant.  Since redundant ground instances remain redundant during saturation, some clauses become eventually redundant in the standard sense, while others that have some redundant ground instances, become \emph{partially redundant}.
}


To capture this idea, we introduce the concept of \emph{redundancy formula}, which can be attached to a clause $C$ and express the fact that all ground instances of $C$ that satisfy a redundancy formula $R$ are redundant. A \emph{partial clause} $C \con R$ is a clause with a redundancy formula attached to it. Our technique allows one to reduce the search space in the following two ways; neither of these steps can be done using the standard methods:

\begin{enumerate}
    \item We can change redundancy formulas to weaker ones to make a clause eventually redundant. For example, we can start with a clause $C$, change it to $C \con {R_1}$, then to $C \con {R_1 \lor R_2}$, etc. If at any step we have a redundancy formula $R_1 \lor \ldots \lor R_n$, which is a tautology, the clause $C$ is redundant.\\
    
    \item Suppose that we apply an inference to a partial clause $C \con R$, which requires an application of a substitution $\sigma$ to $C$. If $R\sigma$ is a tautology, then all instances of this inference are redundant and the inference does not have to be performed at all.
\end{enumerate}


\paragraph{Motivating example.} Let us illustrate our work with the following example. Consider binary function symbols $f$ and $h$, a unary function symbol  $g$, and constants $a$ and $b$. Let  $\succ$ denote a Knuth-Bendix simplification order~\cite{KBO} with constant weight function and precedence $f\gg g\gg h$. Let $C$ be the clause $f(g(y),y)\eqs g(h(x,y))$. The sides of the equation in $C$ are incomparable w.r.t. $\succ$, so standard superposition reasoning has to consider both orientations of this equation. For example, for an arbitrary substitution $\sigma$ that grounds $C$, superposition orients the equation in $C$ right-to-left only if $x\sigma\succ y\sigma$ holds. Now, consider the clause $h(z,u)\eqs h(u,z)$ and the following inference:
\begin{equation*}
\AxiomC{\ensuremath{h(z,u)\eqs h(u,z)}}
\AxiomC{\ensuremath{f(g(y),y)\eqs g(h(x,y))}}
\BinaryInfC{\ensuremath{f(g(y),y)\eqs g(h(y,x))}}
\DisplayProof
\end{equation*}
We can easily show that in the presence of $h(z,u)\eqs h(u,z)$ and $f(g(y),y)\eqs g(h(y,x))$, the right-to-left orientation of the equation in $C$ is not needed anymore, since all instances $C\sigma$ s.t. $x\sigma\succ y\sigma$ are now redundant. In this case, we replace $C$ with the partial clause $C\con x\succ y$ in the search space. As the right-to-left orientation of the equation in $C$ is not needed anymore (item~2 above), we can simply remove the right-hand side of this equation from all term indices, avoiding potentially thousands of unnecessary inferences.\QED

\paragraph{Contributions.} We extend superposition with \emph{partial redundancy} by attaching redundancy formulas to clauses. Further, 
\begin{enumerate}[itemsep=5pt,leftmargin=1.4em]
\item We introduce the \emph{partial redundancy calculus} \Calc\ using partial clauses and redundancy formulas (Section~\ref{sec:calculus}). We prove soundness and refutational completeness in Sections~\ref{sec:saturation}--\ref{sec:model-construction}.
\item We demonstrate the strength of \Calc{} by introducing a generalization of demodulation (Section~\ref{sec:examples}). We show that \Calc{} significantly reduces the number of superposition inferences performed during proof search. 
\item We implemented \Calc{} in \vampire{} and show that we could solve very hard problems that no other prover was able to solve before (Section~\ref{sec:vampire}).
\end{enumerate}


\section{Preliminaries}

We work in standard \emph{first-order logic with equality}, where equality is denoted by $\eqs$. We use variables $x$, $y$, $z$, $u$, $v$, $w$ and terms $s$, $t$, $l$, $r$, all possibly with indices.
A literal is an unordered pair of terms with polarity, i.e., an equality $s\eqs t$ or a disequality $s\neqs t$. We write $s\deqs t$ for either an equality or a disequality. A \emph{clause} is a multiset of literals. We denote clauses by $B, C, D$ and reserve  $\square$ for the \emph{empty clause} that is logically equivalent to $\bot$.

An \textit{expression $E$} is a term, literal or clause. We consider as expressions also redundancy formulas and partial clauses introduced later (Section~\ref{sec:calculus}). An expression is called \emph{ground} if it contains no variables. The set of all ground instances of an expression $E$ is denoted $E^*$. We write $E[s]$ to state that the expression $E$ contains a distinguished occurrence of the term $s$ at some position.  Further, $E[s\mapsto t]$  denotes that this occurrence of $s$ is replaced with $t$; when $s$ is clear from the context, we simply write $E[t]$. We say that $t$ is a \emph{subterm} of $s[t]$, denoted by $t\trianglelefteq s[t]$;  and a \emph{strict subterm} if additionally $t\neq s[t]$, denoted by $t\triangleleft s[t]$.

A \emph{substitution} $\sigma$ is a mapping from variables to terms, such that the set of variables $\{x\mid \sigma(x)\neq x\}$ is finite. We denote substitutions by $\theta$, $\sigma$, $\rho$, $\mu$, $\eta$. The application of a substitution $\theta$ on an expression $E$ is denoted $E\theta$. A substitution $\theta$ is called \emph{grounding for an expression $E$} if $E\theta$ is ground. A substitution $\theta$ is \emph{more general} than a substitution $\sigma$ if $\theta\eta=\sigma$ for some substitution $\eta$. A substitution $\theta$ is a \emph{unifier} of two terms $s$ and $t$ if $s\theta= t\theta$, and is a \emph{most general unifier (mgu)}, denoted $\mgu(s,t)$, if for every unifier $\eta$ of $s$ and $t$, there exists a substitution $\mu$ s.t. $\eta=\theta\mu$. We assume that the mgus are idempotent~\cite{TermRewritingAndAllThat_EquationalProblems}.

A binary relation $\to$ over the set of terms is a \emph{rewrite relation} if (i) $l\to r \Rightarrow l\theta\to r\theta$ and (ii) $l\to r \Rightarrow s[l]\to s[r]$ for any term $l$, $r$, $s$ and substitution $\theta$. We write $\leftarrow$ to denote the inverse of $\to$. We call an ordered pair $l\to r$ a \emph{rewrite rule} if (i) $l$ is not a variable and (ii) $l$ contains all variables that occur in $r$. A \emph{rewrite system} $R$ is a set of rewrite rules. We denote by $\to_R$ the smallest rewrite relation that contains $R$. A term $l$ is \emph{irreducible} in $R$ if there is no $r$ s.t. $l\to r \in R$. A \emph{rewrite ordering} is a strict (irreflexive) rewrite relation. A \emph{reduction ordering} is a well-founded rewrite ordering. We consider reduction orderings which are total on ground terms, that is they satisfy $s \triangleright t \Rightarrow s \succ t$; such orderings are also  called \emph{simplification orders}. 

We use the standard definition of a \emph{bag extension} of an ordering~\cite{MultisetOrderings}. An ordering $\succ$ on terms induces an ordering on literals by identifying $s\eqs t$ with the multiset $\{s,t\}$ and $s\neqs t$ with the multiset $\{s,s,t,t\}$, and using the bag extension of $\succ$. We denote this induced ordering on literals also with $\succ$. Likewise, the ordering $\succ$ on literals induces the ordering on clauses by using the bag extension of $\succ$. Again, we denote this induced ordering on clauses also with $\succ$.  The induced relations $\succ$ on literals and clauses are well-founded (resp. total) if the original relation $\succ$ on terms is too. In examples used in this paper, we assume a Knuth Bendix order (KBO) with constant weight~\cite{KBO}.

Many first-order theorem provers work with clauses~\cite{E19,iProver,CAV13}. Let $S$ be a set of clauses. An inference system $\mathcal{I}$ is a set of inference rules of the form
\vspace{-.5em}
\begin{prooftree}
\def\labelSpacing{1pt}
\AxiomC{$C_1$}
\AxiomC{$\ldots$}
\AxiomC{$C_n$}
\RightLabel{,}
\TrinaryInfC{$C$}
\end{prooftree}
where $C_1,\ldots, C_n$ are the \emph{premises} and $C$ is the \emph{conclusion} of the inference. 
We also use the notation $C_1,\ldots,C_n\infers C$ for inferences.
The inference rule is \emph{sound} if its conclusion is the logical consequence of its premises, that is $C_1,\ldots, C_n\models C$. An inference system $\mathcal{I}$ is \emph{sound} if all its inferences are sound. An inference system $\mathcal{I}$ is \emph{refutationally complete} if, for every unsatisfiable clause set $S$, there is a derivation of the empty clause in $\mathcal{I}$.

\section{The Partial Redundancy Calculus}\label{sec:calculus}
We introduce the notion of \emph{partial redundancy} and describe a general calculus, called \Calc{}, based on it. Soundness and completeness of \Calc{}  are proved in Sections~\ref{sec:saturation}--\ref{sec:model-construction}.

Let us first define the key notions of redundancy formulas and partial clauses.
We assume to have a finite signature $\Sigma$ consisting of function symbols and having at least one constant. As usual, the \emph{term algebra} over $\Sigma$, denoted by $T(\Sigma)$, is the algebra with the universe consisting of the ground terms built from symbols in $\Sigma$ and having $\Sigma$ as the signature, such that the interpretation of every ground term $t$ in $T(\Sigma)$ is $t$ itself. We will consider \emph{extensions of $T(\Sigma)$ by a finite set $\mathcal{R}$ of predicate symbols}, each one having a fixed interpretation. We will denote these extensions by $T_\mathcal{R}(\Sigma)$ and have  their interpretations  as \emph{redundancy relations}.

Let us assume that we have a simplification order $\succ$ on terms. We consider $\succ$ to be a redundancy relation and assume that $\mathcal{R}$ contains the symbol $\succ$ interpreted as this simplification order. In general, we can also consider redundancy relations different from orderings. For the moment, we will leave the choice of suitable redundancy relations open. In a way, $T_\mathcal{R}(\Sigma)$ is a Herbrand interpretation, with the difference that relations in it are not taken from clauses.
\begin{Definition}[Redundancy Formula and Partial Clause]
  A \emph{redundancy formula} $\rf$ is an arbitrary first-order formula of $T_\mathcal{R}(\Sigma)$.
  A \textit{partial clause} is an expression $C\con{\rf}$, where $C$ is a clause and $\rf$ a redundancy formula
  whose free variables occur in $C$. If $\sigma$ is a substitution and $R$ a redundancy formula, we write $\sigma \vDash R$ if every ground instance of $R\sigma$ is true in $T_\mathcal{R}(\Sigma)$. We write $\sigma \nvDash R$ if $\sigma \vDash R$ does not hold.
  \QED
\end{Definition}
An example of a redundancy formula is $f(x) \succ y \lor y = x$. Note that we can use equality $=$ in redundancy formulas as in this example.
In the sequel, we denote redundancy formulas by $\rf$, and  partial clauses by $\CC$ and $\CD$, possibly with indices. We  sometimes write $C$ instead of $C\con{\bot}$. 
Logically, a partial clause $C\con{\rf}$ is the same as $C$. The intuitive semantics of partial clauses comes from their use in saturation: when a partial clause $C\con{\rf}$ occurs in a search space, we know that all ground instances of $C$ that satisfy $\rf$ are redundant in this search space.

 Let us introduce some abbreviations for redundancy formulas. Let $s$, $t$ and $l$ be terms. We will write $s \neq t$ for $\lnot(s = t)$; $s \succeq t$ for $s \succ t \lor s = t$. If $s \deqs t$ is an equality literal, we write $(s \deqs t) \succ l$ for $s \succ l \lor t \succ l$ and $(s \deqs t) \succeq l$ for $s \succeq l \lor t \succeq l$. If $C$ is a clause $L_1 \lor \ldots \lor L_n$, we write $C \succ l$ for $L_1 \succ l \lor \ldots \lor L_n \succ l$, and $C \succeq l$ for $L_1 \succeq l \lor \ldots \lor L_n \succeq l$.
We assume a literal selection function satisfying the standard condition on $\succ$ and underline selected literals. 

Our Partial Redundancy Calculus (\Calc{}) dealing with partial clauses is displayed in Figure~\ref{fig:calculus}.

\begin{figure}[t]
\normalsize
\begin{tabular}{c >{\centering}m{.08\linewidth} l}
\multirow{8}{*}{
\AxiomC{$\underline{l\eqs r}\lor C_1\con{\rf_1}$}
\AxiomC{$\underline{s[l']\deqs t}\lor C_2\con{\rf_2}$}
\LeftLabel{($\Sup$)}
\BinaryInfC{$(s[r]\deqs t\lor C_1 \lor C_2)\sigma$}
\DisplayProof
}
& \multirow{8}{*}{where} & (1) $\sigma=\mgu(l,l')$\\
& & (2) $l'$ is not a variable\\
& & (3) $\sigma \nvDash r \succeq l$\\
& & (4) $\sigma \nvDash t  \succeq s[l']$ \\
& & (5) $\sigma \nvDash C_1 \succeq l$\\
& & (6) $\sigma \nvDash C_2 \succeq s[l']$ (if\\
& & \phantom{(6) }$s[l']\deqs t$ is positive)\\
& & (7) $\sigma \nvDash R_1$\\
& & (8) $\sigma \nvDash R_2$ \\[1em]

\multirow{2}{*}{
\AxiomC{$\underline{s\neqs t}\lor C \con{\rf}$}
\LeftLabel{($\EqRes$)}
\UnaryInfC{$C\sigma$}
\DisplayProof
}
& \multirow{2}{*}{where} & (1) $\sigma=\mgu(s,t)$\\
& & (2) $\sigma\nvDash R$\\[1em]

\multirow{5}{*}{
\AxiomC{$\underline{s\eqs t}\lor \underline{s'\eqs t'} \lor C\con{\rf}$}
\LeftLabel{($\EqFac$)}
\UnaryInfC{$(s\eqs t\lor t\neqs t' \lor C)\sigma$}
\DisplayProof
}
& \multirow{5}{*}{where} & (1) $\sigma=\mgu(s,s')$\\
& & (2) $\sigma \nvDash t \succeq s$\\
& & (3) $\sigma \nvDash t' \succ t$\\
& & (4) $\sigma \nvDash C \succ s$\\
& & (5) $\sigma\nvDash R$\\[1em]
\end{tabular}
\vspace{-1em}
\caption{The \Calc{} calculus.}
\vspace{-1em}
\label{fig:calculus}
\end{figure}
\section{Saturation with \Calc{}}
\label{sec:saturation}

This section  explains how saturation algorithms work on partial clauses and concludes completeness of \Calc. We assume a fixed
signature $\Sigma$.

\begin{Definition}[Redundant Clause]
A ground clause $D$ is \emph{redundant w.r.t.\ a set of clauses} $S$, 
if there are clauses $C_1,\ldots,C_n\in S$ s.t.\ $C_1,\ldots,C_n\vDash D$ and $D \succ C_i$ for all $1\leq i\leq n$.\QED
\end{Definition}
Let $C$ be a clause. Then $C^*$ denotes the set of all ground instances of $C$. For a set $S$ of clauses, the set $S^*$ is defined as $\bigcup_{C\in S} C^*$. 

\begin{Definition}[Saturation]
\label{def:saturation}
A \emph{saturation} $\satS$ is a sequence of sets of partial clauses $S_0,S_1,\ldots,$ such that $S_0$ is a set of partial clauses of the form $C\con\bot$, and for each  $i$, we have one of the following cases:
\begin{enumerate}[itemsep=5pt,leftmargin=1.4em]
    \item $S_{i + 1}$ is obtained by adding to $S_i$ the conclusion of a \Calc{} inference with premises in $S_i$. We then say that we \emph{apply this inference at the step} $S_{i+1}$.
    \item $S_{i+1}$ is obtained from $S_i$ by changing a partial clause $C \con{R_1}$ to $C \con{R_1 \lor R_2}$, such that for every substitution $\theta$ grounding for $C$, if $\theta \vDash R_2$, then $C\theta$ is redundant w.r.t.\ $S^*_i$. 
\end{enumerate}
We call clauses in $S_0$ the \emph{initial clauses} of this saturation. We  also refer to steps which apply an inference as \emph{inference steps} and steps that update redundancy formulas as \emph{redundancy steps}. \QED
\end{Definition}
To prove completeness of \Calc{}, we introduce notions of fairness and persistent ground clauses, which are modifications of similar notions used in completeness proofs of superposition calculi~\cite{NieuwenhuisRubio:HandbookAR:paramodulation:2001}. In the sequel, we assume that $\satS$ is a saturation $S_0,S_1,\ldots$ and denote by $S_\omega$ the set $\bigcup_{i \geq 0}S_i$. Let $C \con R$ be a partial clause and $S$ a set of partial clauses. We define
\[
\begin{array}{rcl}
     \gnr{(C \con R)} & \defeq & \setof{C \theta \mid C\theta \text{ is ground and } \theta \nvDash R};  \\*[1ex]
     \gnr{S} & \defeq & \bigcup_{\CC \in S} \gnr{\CC}.
\end{array}
\]

\begin{Definition}[Limit and Persistent Clause]
The \emph{limit} of a saturation $\satS$ is the set of ground clauses $\bigcup_{i \geq 0}\bigcap_{j \geq i}\gnrS_j$. We call a ground clause $C$ \emph{persistent} if $C$ belongs to the limit.\QED
\end{Definition}
It is not hard to argue that a ground clause $C$ is persistent if and only if there exists $i \geq 0$ such that for all $j \geq i$ we have $C \in \gnrS_j$.

\begin{Definition}[Fair Saturation]
A saturation $\satS$ is \emph{fair} if it has the following property.
Suppose that 
\begin{enumerate}
    \item $D_1,\ldots,D_n$ are persistent ground clauses;
    \item there exists an inference $C_1 \con {R_1},\ldots,C_n \con{R_n} \infers C\sigma$ and a substitution $\rho$ such that for all $i = 1,\ldots, n$ we have that $D_i = C_i\sigma\rho$.
\end{enumerate}
Then $\satS$ contains a step $i$ such that $C\sigma\con\bot\in S_i$.\QED
\end{Definition}
Note that our notion of fairness is not a standard one: it reflects the fact that the redundancy formula attached to a clause $C$ may change when we apply redundancy steps. 
Indeed, we can have a partial clause $C \con{R_1}$ that can be replaced by $C \con{R_1 \lor R_2}$, then replaced by $C \con{R_1 \lor R_2 \lor R_3}$, and so on. 
Our notion of fairness requires that we apply the rule to \emph{some} of these partial clauses, but not necessarily to $C \con{R_1}$.

Our goal is to prove  soundness and completeness of the \Calc{} calculus.
\begin{theorem}[Completeness\label{thm:parc:compl}]\rm
Let $\satS$ be a fair saturation $S_0, S_1, \ldots$. Then, $S_0$ is unsatisfiable if and only if  the empty partial clause $\emptyclause \con \bot$ occurs in $S_i$ for some~$i$.
\end{theorem}
We will not give a full proof of Theorem~\ref{thm:parc:compl}, as it repeats standard completeness proof arguments. Instead, we  give the sketch of the proof and consider a few representative cases in detail (Section~\ref{sec:model-construction}), to illustrate how we deal with redundancy formulas and a slightly non-standard notion of fairness.\cameraready{\footnote{Detailed proofs can be found in the extended version of this paper~\cite{FullPaper}.}}


\section{Model Construction}\label{sec:model-construction}
We starts this section by proving soundness of \Calc{}.
\begin{Definition}
Let $C$ be a clause and $S$ a set of partial clauses. We say that \emph{$C$ occurs in $S$}, or that \emph{$S$ contains $C$} (denoted by $C \dotin S$) if there exists a redundancy formula $R$ such that $(C \con R) \in S$.\QED
\end{Definition}
\begin{restatable}[Soundness]{Lemma}{soundnessLemma}
    Let $C$ be a clause such that $C \dotin S_\omega$. Then $C$ is a logical consequence of clauses in $S_0$.
    Hence, if $\emptyclause \dotin S_\omega$, then $S_0$ is unsatisfiable.
\end{restatable}
\fullversion{
\begin{proof}
By induction on $i$. Note that inference steps preserve this property and  redundancy steps do not change the set of clauses occurring in $S_i$. \QED    
\end{proof}
}
\noindent
To prove completeness (Theorem~\ref{thm:parc:compl}), we  introduce several properties.
\begin{restatable}{Lemma}{redundancyLemma}\label{lem:redundancy}%
If a ground clause $C$ is redundant with respect to $S^*_i$, then for all $j \geq i$, this clause is  redundant with respect to $S^*_j$ and also with respect to $\gnrS_j$.
\end{restatable}
\fullversion{
\begin{proof}
Standard using well-foundedness of $\succ$.\QED
\end{proof}
}
\noindent
Note that, as a consequence, a ground clause is redundant w.r.t.\ $S^*_j$ if and only if it is redundant with respect to $\gnrS_j$, for all $j = 0,1,\ldots$.
\begin{Definition}[Trace]%
The \emph{trace} of a ground clause $C$ is any finite set $\{C_1$, $\ldots$, $C_n\}$ of persistent clauses such that $C_1,\ldots,C_n \vDash C$ and, for all $i = 1,\ldots, n$, we have $C \succeq C_i$.\QED
\end{Definition}
Let us prove that every ground instance of every clause in $S_\omega$ has a trace.
\begin{restatable}{Lemma}{traceLemma}\label{lem:trace}%
Suppose that $C \dotin S_\omega$ and $\theta$ is a substitution grounding for $C$. Then $C\theta$ has a trace.
\end{restatable}
\fullversion{
\begin{proof}
    We prove it by induction on $\succ$ over clauses.
    If $C\theta$ is persistent, then $\{C\theta\}$ is its trace. 
    Suppose that $C\theta$ is not persistent.
    This can only happen if for some $i=0,1\ldots$ and some redundancy formula $R$ we have that $(C \con R) \in S_i$ and $\theta \vDash R$. 
    Take the smallest $i$ with this property. 
    
    Note that by $\theta \vDash R$ we have that $R$ is different from $\bot$. 
    Therefore, by the definition of saturation, $i$ is a redundancy step changing $C \con R_1$ to $C \con R_1 \lor R_2$, and we have that $\theta \vDash R_2$ and $C\theta$ is redundant w.r.t.\ $S_i^*$. 
    Then there exist ground clauses $C_1,\ldots,C_n$ in $S^*_i$ such that $C_1,\ldots,C_n \vDash C\theta$ and for all $1\le j\le n$ we have $C\theta \succ C_j$. 
    By induction, each of the clauses $C_j$ has a trace. It is not hard to argue that the union of traces of all $C_j$ is a trace of $C\theta$. \QED
\end{proof}
}

\begin{restatable}{Lemma}{RMeansRedundant}\label{lem:R:means:redundant}%
    Suppose that $(C \con R) \in S_i$, $\theta$ is a substitution grounding for $C$ and $\theta \vDash R$. Then $C\theta$ is redundant w.r.t.\ $S_i^*$. 
\end{restatable}
\fullversion{
\begin{proof}
    Similar to the proof of Lemma~\ref{lem:trace}.\QED
\end{proof}
}

%
%
%

Now, as usual in completeness proofs, we will build an interpretation for the set of partial clauses $S_\omega$ by well-founded induction on the relation $\succ$ on terms and prove that this interpretation satisfies all persistent ground clauses.

We assume familiarity with standard notions of rewrite systems.
Let $I$ be a convergent ground rewrite system on terms of the signature $\Sigma$. We can consider $I$ also as a first-order interpretation (or a factor-algebra of $T(\Sigma)$) by defining $I \vDash s\eqs t$ if and only if $s$ and $t$ have the same normal form in $I$.

To define the model of $S_\omega$, for each ground term $l$, we  define two convergent rewrite systems $I_{l}$ and $I_{\prec l}$. First, we set 

  \[
     I_{\prec l}\ \defeq\ \bigcup_{\setof{r \mid l \succ r}} I_r. 
  \]
We next define $I_l$ using $I_{\prec l}$. Suppose that $l$ is irreducible in $I_{\prec l}$ and there exist a partial clause $(\underline{l_1 \eqs r_1} \lor C \con{R}) \in S_\omega$ and a substitution $\theta$ grounding for this clause such that
\begin{enumerate}
    \item $l_1\theta = l$;
    \item $l_1\theta\succ r_1\theta$;
    \item $l_1\theta\succ C\theta$;
    \item $\theta \nvDash R$;
    \item $I_{\prec l} \nvDash C\theta$.
\end{enumerate}
Then, we let $I_l \defeq I_{\prec l} \cup \setof{l \rightarrow r_1\theta}$. 
If there is more than one partial clause $\underline{l_1 \eqs r_1} \lor C \con{R}$ and substitution $\theta$ with this property, we can select an arbitrary one.
In all other cases we let $I_l \defeq I_{\prec l}$. Finally, we define 

\[
  I_\omega\ \defeq\ \bigcup_{l} I_l.
\]
Let us note some properties of this construction. They are used in many completeness proofs of variants of the superposition calculus (and are straightforward), so we will not prove them. In these properties, $l,s,t$ are ground terms and $C$ is a ground clause:

\begin{enumerate}
    \item $I_\omega$ is non-overlapping and terminating, and so it is convergent.
    \item Let $s \succ t$. Suppose that for every term $r$ occurring in $C$, we have $t \succeq r$. Then $I_s \vDash C$ if and only if $I_t \vDash C$. 
    \item Suppose that for every term $r$ occurring in $C$, we have $l \succeq r$. Then $I_\omega \vDash C$ if and only if $I_l \vDash C$. 
    \item Let $s \succ t$. Then $t$ is irreducible in $I_t$ 
    if and only if $t$ is irreducible in $I_s$ and if and only if $t$ is irreducible in $I_\omega$.
    \item If $s \succ t$ and $s$ is irreducible w.r.t.\ $I_\omega$, then $I_\omega \vDash s \neqs t$.
\end{enumerate}
Let us now give a useful property of persistent clauses.
\begin{restatable}{Lemma}{falsePersistentLemma}\label{lem:false:persistent}%
    Suppose that $C$ is a ground clause, $C \in S_i^*$ and $I_\omega \nvDash C$. Then there exists a persistent ground clause $D$ such that $C \succeq D$ and $I_\omega \nvDash D$.
\end{restatable}
\fullversion{
\begin{proof}
    By Lemma~\ref{lem:trace}, $C$ has a trace, that is, there are persistent ground clauses $C_1,\ldots,C_n$ in $S_i^*$ such that $C_1,\ldots,C_n \vDash C$ and for all $j = 1,\ldots,n$ we have $C\succeq C_j$.
    Since $I_\omega \nvDash C$, then there is $j$ such that $I_\omega \nvDash C_j$. 
    Note that $C_j$ is persistent and $C \succeq C_j$, so we are done.\QED
\end{proof}
}
\noindent
We are now ready to conclude completeness of \Calc{} with the following lemma.
\begin{restatable}{Lemma}{mainLemma}
    For every persistent ground clause $B$, we have $I_\omega \vDash B$.
\end{restatable}

\review{
\begin{proof}
We prove by induction on the ordering $\succ$ on clauses. By contradiction, suppose that this is not the case, then there is a persistent clause false in $I_{\omega}$. Since $\succ$ is well-founded, then there is a persistent clause $B$ that is minimal among persistent clauses false in $I_{\omega}$.

We prove that this is impossible by finding a persistent clause that is false in $I_{\omega}$ and smaller than $B$. \cameraready{Here, we only prove a few representative cases. The full proof can be found in the extended version of this paper~\cite{FullPaper}.} \medskip

\noindent
\begin{Case}{1}{There is a step $i$, a partial clause $C\con{\rf}\in S_i$ and a ground substitution $\theta$ such that $C\theta=B$ and $\theta\vDash R$.}

\noindent
By Lemma~\ref{lem:R:means:redundant}, we have that $B$ is redundant in $S_i^*$.
By Lemma~\ref{lem:false:persistent}, there exists a ground persistent clause $D$ such that $B \succeq D$ and $I_\omega \nvDash D$. Note that $B$ is different from $D$ since $B$ is redundant, hence we have $B \succ D$.
Thus, we have found a persistent false clause smaller than $B$.
\end{Case}

\begin{Case}{2}{$B$ has the form $\underline{s' \neqs s'} \lor C'$.}

\noindent
Note that in this case, we also have $I_\omega \nvDash C'$. Since $B$ is a ground instance of a clause occurring in the saturation, there exists a step $i$, a partial clause $(\underline{s \neqs t} \lor C \con R) \in S_i$, and a ground substitution $\theta$ such that $s\theta = t\theta = s'$ and $C\theta = C'$. Note that $s$ and $t$ are unifiable. Let $\sigma = \mgu(s,t)$. Since the saturation is fair, it has a step $j$, where the following rule is applied:

\begin{prooftree}
\AxiomC{$\underline{s\neqs t}\lor C \con{\rf'}$}
\LeftLabel{$\phantom{(\EqRes)}$}
\RightLabel{$(\EqRes)$}
\UnaryInfC{$C\sigma$}
\end{prooftree}
Note that by $\theta\nvDash R'$ (since Case 1 does not apply), we also have $\sigma\nvDash R'$ and side condition (2) of \EqRes{} is satisfied. We have that $(C\sigma \con\bot) \in S_j$. But $C'$ is a ground instance of $C\sigma$, so $C' \in S^*_j$. Using $I_\omega \nvDash C'$ and Lemma~\ref{lem:false:persistent}, there exists a persistent ground clause $D$ such that $C' \succeq D$ and $I_\omega \nvDash D$. Note that we have $B \succ D$, so we have found a persistent false clause smaller than $B$.
\end{Case}

\begin{Case}{3}{$B$ has the form $\underline{s' \neqs t'} \lor C'$.}

\noindent
W.l.o.g. $s'\succ t'$. Since $B$ is a ground instance of a clause occurring in the saturation, there exists a step $i$, a partial clause $(s \neqs t \lor C \con R_2) \in S_i$, and a ground substitution $\theta_2$ such that $s\theta_2 = s'$, $t\theta_2 = t'$ and $C\theta_2 = C'$.\medskip

\begin{Subcase}{3.1}{$s'$ is irreducible in $I_\omega$.}

\noindent
In this case, $I_\omega \vDash s' \neqs t'$ contradicts the assumption that $I_\omega \nvDash B$.
\end{Subcase}

\begin{Subcase}{3.2}{$s'$ is reducible in $I_\omega$.}

\noindent
In this case, $s'$ contains a subterm $l'$ such that for some ground term $r'$, we have $(l' \rightarrow r') \in I_\omega$. 
Now $B$ has the form $\underline{s'[l'] \neqs t'} \lor C'$. 
Since $(l' \rightarrow r') \in I_\omega$, by construction we have that $l'$ is irreducible in $I_{\prec l'}$ and there exist a partial clause $(\underline{l_1 \eqs r_1} \lor C_1 \con{R_1}) \in S_\omega$ and a substitution $\theta_1$ grounding for this clause such that $l_1\theta_1 = l'$, $l_1\theta_1\succ r_1\theta_1$, $l_1\theta_1\succ C_1\theta_1$, $\theta_1 \nvDash R_1$ and $I_{\prec l'} \nvDash C_1\theta_1$.

As usual, we assume that clauses have pairwise disjoint sets of variables and we can define a substitution $\theta$ as a union of $\theta_1$ and $\theta_2$. Then $\theta$ will satisfy all properties we used for $\theta_1$ and $\theta_2$ above. 

Suppose that there is a variable $x$ in $s$ such that $x\theta$ contains an occurrence of $l'$. Let $\theta'$ be a substitution identical to $\theta$ except that it maps $x$ to the normal form of $x\theta$ in $I_\omega$. Since for all variables $y$ we have $y\theta\succeq y\theta'$ and $x\theta\succ x\theta'$, we also have $B=(s\neqs t\lor C)\theta\succ (s\neqs t\lor C)\theta'$. Note that $(s\neqs t\lor C)\theta$ and $(s\neqs t\lor C)\theta'$ are equivalent in $I_\omega$, so $I_\omega \nvDash (s\neqs t\lor C)\theta'$. By Lemma~\ref{lem:false:persistent}, there is a persistent ground clause $D$ such that $(s\neqs t\lor C)\theta' \succeq D$ and $I_\omega \nvDash D$. But then we also have $(s\neqs t\lor C)\theta \succ D$, and therefore we have found a false persistent clause smaller than $B$.

Otherwise, there must be a subterm $l$ in $s$ such that $l\theta=l'$. Note that $l$ and $l_1$ are unifiable. Let $\sigma = \mgu(l,l_1)$. Since the saturation is fair, it has a step $j$, where the following rule is applied:

\begin{prooftree}
\AxiomC{$\underline{l_1\eqs r_1}\lor C_1\con{\rf'_1}$}
\AxiomC{$\underline{s[l]\neqs t}\lor C\con{\rf'_2}$}
\LeftLabel{$\phantom{(\Sup)}$}
\RightLabel{$(\Sup)$}
\BinaryInfC{$(s[r_1]\neqs t\lor C_1 \lor C)\sigma$}
\end{prooftree}
Conditions (3) and (5) of \Sup{} are satisfied due to $l_1\theta_1\succ r_1\theta_1$ and $l_1\theta_1\succ C_1\theta_1$. Condition (4) of \Sup{} is satisfied due to $s'\succ t'$. By assumption, we have $\theta\nvDash R_1'$ and $\theta\nvDash R_2'$, satisfying conditions (7) and (8) of \Sup{}.

We have that $((s[r_1]\neqs t\lor C_1 \lor C)\sigma \con\bot) \in S_j$. The ground clause $s'[r']\neqs t'\lor C'_1 \lor C'$ is an instance of $(s[r_1]\neqs t\lor C_1 \lor C)\sigma$, so we have $(s'[r']\neqs t'\lor C'_1 \lor C') \in S^*_j$. It is also not hard to argue that $I_\omega \nvDash s'[r']\neqs t'\lor C'_1 \lor C'$. Using this and Lemma~\ref{lem:false:persistent}, there exists a persistent ground clause $D$ such that $(s'[r']\neqs t'\lor C'_1 \lor C') \succeq D$ and $I_\omega \nvDash D$. Finally, we note that $B \succ (s'[r']\neqs t'\lor C'_1 \lor C')$, hence we have $B \succ D$, so we have found a persistent false clause smaller than $B$.
\end{Subcase}
\end{Case}
\cameraready{\QED}
\fullversion{

\begin{Case}{4}{$B$ has the form $\underline{s'\eqs t'}\lor C'$.}

W.l.o.g., we can assume that $s'\succ t'$ and that $s'\eqs t'$ is maximal in $B$ by literal selection. Since $B$ is a ground instance of a clause occurring in the saturation, there exists a step $i$, a partial clause $(\underline{s\eqs t}\lor C\con R)\in S_i$, and a ground substitution $\theta$ such that $s\theta=s'$, $t\theta=t'$ and $C\theta=C'$.

\begin{Subcase}{4.1}{$s'$ is irreducible in $I_\omega$.}

Then, it must be the case that $s'\nsucc C'$, which means that $B$ is of the form $\underline{s'\eqs t'}\lor s'\eqs r'\lor C''$ and the partial clause $\underline{s\eqs t}\lor C\con R$ is of the form $\underline{s\eqs t}\lor l\eqs r\lor C\con R$ where $l\theta=s'$ and $r\theta=r'$. Let $\sigma=\mgu(s,l)$. Since the saturation is fair, it has a step $j$, where the following rule is applied:

\begin{prooftree}
\AxiomC{$\underline{s\eqs t}\lor l\eqs r \lor C\con{\rf'}$}
\LeftLabel{\phantom{($\EqFac$)}}
\RightLabel{($\EqFac$)}
\UnaryInfC{$(s\eqs t\lor t\neqs r \lor C)\sigma$}
\end{prooftree}

Condition (2) of \EqFac{} is satisfied due to $s'\succ t'$, and conditions (3) and (4) of \EqFac{} are satisfied due to $s'\succ t'$ and $s'\eqs t'$ being maximal. Finally, condition (5) of \EqFac{} is also satisfied due to $\theta\nvDash R'$.

We have that $((s\eqs t\lor t\neqs r\lor C)\sigma\con\bot)\in S_j$. The ground clause $s'\eqs t'\lor t'\neqs r'\lor C'$ is an instance of $(s\eqs t\lor t\neqs r\lor C)\sigma$, so we have $s'\eqs t'\lor t'\neqs r'\lor C'\in S_j^*$. It is also not hard to argue that $I_\omega\nvDash s'\eqs t'\lor t'\neqs r'\lor C'$. By Lemma~\ref{lem:false:persistent}, there exists a persistent ground clause $D$ such that $s'\eqs t'\lor t'\neqs r'\lor C'\succeq D$ and $I_\omega\nvDash D$. We note that $B\succ s'\eqs t'\lor t'\neqs r'\lor C'$, so we have found a persistent ground clause that is false and smaller than $B$.
\end{Subcase}

\begin{Subcase}{4.2}{$s'$ is reducible in $I_\omega$.}

This case is similar to Subcase 3.2. Again, $s'$ contains a subterm $l'$ such that for some ground term $r'$ we have $(l' \rightarrow r') \in I_\omega$ and there exists a partial clause $(\underline{l_1\eqs r_1} \lor C_1 \con{R_1}) \in S_\omega$ that has no variables in common with $\underline{s\eqs t}\lor C\con R$ and so $\theta$ can be extended such that it is also grounding for this clause and $l_1\theta = l'$, $l_1\theta\succ r_1\theta$, $l_1\theta\succ C_1\theta$, $\theta \nvDash R_1$ and $I_{\prec l'} \nvDash C_1\theta$.

The case when there is a variable $x$ in $s$ such that $x\theta$ contains an occurrence of $l'$ is the same as in Subcase 3.2. Otherwise, there must be a subterm $l$ in $s$ such that $l\theta=l'$. Note that $l$ and $l_1$ are unifiable. Let $\sigma = \mgu(l,l_1)$. Since the saturation is fair, it has a step $j$, where the following rule is applied:

\begin{prooftree}
\AxiomC{$\underline{l_1\eqs r_1}\lor C_1\con{\rf'_1}$}
\AxiomC{$\underline{s[l]\eqs t}\lor C\con{\rf'_2}$}
\LeftLabel{$\phantom{(\Sup)}$}
\RightLabel{$(\Sup)$}
\BinaryInfC{$(s[r_1]\eqs t\lor C_1 \lor C)\sigma$}
\end{prooftree}

Now condition (6) of \Sup{} is satisfied due to $s'\succ t'$ and $s'\eqs t'$ being maximal in $B$. The other conditions hold similarly as in Subcase 3.2.

We have that $((s[r_1]\eqs t\lor C_1 \lor C)\sigma \con\bot) \in S_j$. The ground clause $s'[r']\eqs t'\lor C'_1 \lor C'$ is an instance of $(s[r_1]\eqs t\lor C_1 \lor C)\sigma$, so we have $(s'[r']\eqs t'\lor C'_1 \lor C') \in S^*_j$. It is also not hard to argue that $I_\omega \nvDash s'[r']\eqs t'\lor C'_1 \lor C'$. Using this and Lemma~\ref{lem:false:persistent}, there exists a persistent ground clause $D$ such that $(s'[r']\eqs t'\lor C'_1 \lor C') \succeq D$ and $I_\omega \nvDash D$. Finally, we note that $B \succ (s'[r']\eqs t'\lor C'_1 \lor C')$, hence we have $B \succ D$, so we have found a persistent false clause smaller than $B$.
\end{Subcase}
\end{Case}

\noindent
We have covered all cases, which proves the claim.\QED}
\end{proof}
}

\section{\Calc{} Demodulation and Examples}
\label{sec:examples}

In this section, we demonstrate how the \Calc{} calculus works, and adapt some standard simplifications.  

\begin{Definition}[Admissible Redundancy Formula]
Let $S$ be a set of partial clauses, $C$ a clause and $R$ a redundancy formula. We say that $R$ is \emph{admissible for $S$ and $C$}, if for every substitution $\sigma$ grounding for $C$, if $\sigma \vDash R$, then $C\sigma$ is redundant w.r.t.\ $S^*$.\QED
\end{Definition}
Note that the definition of redundancy steps in saturation requires $R_2$ to be admissible for $S_i$ and $C$. Let us now introduce a redundancy step that generalizes the standard simplifying inference \emph{demodulation}.

%
\begin{restatable}{Lemma}{demodulationLemma}\label{lem:demodulation}%
Let $S_0, \ldots, S_i$ be a saturation s.t.\ the step $S_i$ applies an inference of the form
\vspace{-1em}
\begin{prooftree}
\AxiomC{$l\eqs r\con{\rf_1}$}
\AxiomC{$C[l']\con{\rf_2}$}
\RightLabel{,}
\BinaryInfC{$C[r]\sigma$}
\end{prooftree}
where $\sigma=\mgu(l,l')$. Consider the following redundancy formula:
\[ 
  R\ \defeq\ \exists \bar{y}.\,\big(l=l'\land l\succ r\land C[l']\succ l\big),
\]
where $\bar{y}$ contains all free variables occurring in $l$ and $r$ but not in $C[l']$. Then $R$ is admissible for $S_i$ and $C[l']$.
\end{restatable}
\fullversion{\begin{proof}
Let $\theta$ be a substitution grounding for $C[l']$ s.t. $\theta\vDash R$. Then, in particular, $l'\theta$ can be unified with $l$ and thus there is a substitution $\rho$ s.t. $\sigma\rho\vDash (l = l'\theta\land l\succ r\land C[l']\theta\succ l)$ due to $\sigma=\mgu(l,l')$. Note that $l\sigma\rho$ is ground and so by $\sigma\rho\vDash l\succ r$, the term $r\sigma\rho$ is also ground. Hence, $(l\eqs r)\sigma\rho$ and $C[r]\sigma\theta$ are ground instances in $S_i^*$ due to $(l\eqs r),C[r]\sigma\in S_i$, and they are smaller than and imply $C[l']\theta$. This means that $C[l']\theta$ is redundant w.r.t. $S_i^*$.\QED
\end{proof}}
\noindent
We call a redundancy step that adds the redundancy formula $R$ to $C[l']\con{R_2}$ from \Cref{lem:demodulation} a \emph{demodulation step}. \review{Note that the first two conjuncts $l=l'$ and $l\succ r$ in $R$ ensure that the existential quantifier can be eliminated.}
%
In our examples, we will use binary symbols $f$ and $h$, a unary symbol $g$, constants $a$ and $b$, and assume that $\succ$ is a KBO with precedence $f\gg g\gg h$ and a constant weight function. 

Let us first illustrate a saturation simulating a standard demodulation inference with a demodulation step.
The first example shows a saturation that results in a partially redundant clause.
\begin{example}
We show a saturation below, where each box contains clauses from a saturation step.
\begin{center}
\small
\def\arraystretch{1.2}
\setlength{\abovedisplayskip}{0pt}
\setlength{\belowdisplayskip}{0pt}
\setlength{\abovedisplayshortskip}{0pt}
\setlength{\belowdisplayshortskip}{0pt}
\begin{tabular}{|>{\centering\arraybackslash}p{.3\linewidth}|>{\centering\arraybackslash}p{.3\linewidth}|>{\centering\arraybackslash}p{.35\linewidth}|}
\hline
{\normalsize $S_0$} & {\normalsize $S_1$} & {\normalsize $S_2$}
\\\hline

{\begin{align*}
1.&&f(a,b)&\eqs b\\
2.&&f(a,y)&\eqs g(y)\\
3.&&g(f(x,b))&\neqs a\\
\end{align*}}
&

{\begin{align*}
1.&&f(a,b)&\eqs b\\
2.&&f(a,y)&\eqs g(y)\\
3.&&g(f(x,b))&\neqs a\\
4.&&g(g(b))&\neqs a
\end{align*}}
&

{\begin{align*}
1.&&f(a,b)&\eqs b\\
2.&&f(a,y)&\eqs g(y)\\
3.&&g(f(x,b))&\neqs a\con{x=a}\\
4.&&g(g(b))&\neqs a
\end{align*}}

\\\hline

\end{tabular}
\end{center}
We obtain $S_1$ by adding to $S_0$ the conclusion of the following superposition into clause 3 with clause 2:
\vspace{-.5em}
\begin{prooftree}
\AxiomC{$f(a,y)\eqs g(y)$}
\AxiomC{$g(f(x,b))\neqs a$}
\BinaryInfC{$g(g(b))\neqs a$}
\end{prooftree}
We apply a demodulation step, generating the following redundancy formula based on the above inference:
$$\exists y.\,f(a,y) = f(x,b)\land f(a,y)\succ g(y)\land (g(f(x,b))\neqs a)\succ f(a,y)$$
The first conjunct of $R$ simplifies to $a = x\land y = b$, so we can eliminate the quantifier to simplify the formula to:
$$x = a\land f(a,b)\succ g(b)\land (g(f(a,b))\neqs a)\succ f(a,b)$$
The second conjunct is true, and the third conjunct is $g(f(a,b))\succ f(a,b)\lor a\succ f(a,b)$, which is also true. Hence, $S_2$ is a demodulation step adding the redundancy formula $x = a$ to $g(f(x,b))\neqs a$.
Consider the following superposition in $S_2$ with clause 1 into clause 3:
\vspace{-.5em}
\begin{prooftree}
\AxiomC{$f(a,b)\eqs b$}
\AxiomC{$g(f(x,b))\neqs a\con{x = a}$}
\BinaryInfC{$g(b)\neqs a$}
\end{prooftree}
The unifier of the inference is $\sigma=\{x\mapsto a\}$, and the inference is not performed due to condition (4) of \Sup{}, as $\sigma\vDash x=a$. \review{We conclude this example by noting that the clause $g(b)\neqs a$ can still be derived, by superposing with clause 1 into clause 2, and then superposing with the resulting clause into clause 4.}\QED
\end{example}
%
The next example shows that \Calc{} can detect redundant \emph{orientations} of equations.
\begin{example}
\label{ex:redundant-orientation}
Consider the following saturation:
\begin{center}
\small
\def\arraystretch{1.2}
\setlength{\abovedisplayskip}{0pt}
\setlength{\belowdisplayskip}{0pt}
\setlength{\abovedisplayshortskip}{0pt}
\setlength{\belowdisplayshortskip}{0pt}
\begin{tabular}{|>{\centering\arraybackslash}p{.4\linewidth}|>{\centering\arraybackslash}p{.03\linewidth}|>{\centering\arraybackslash}p{.5\linewidth}|}

\cline{0-0}\cline{3-3}
{\normalsize $S_0$} & & {\normalsize $S_2$}
\\\cline{0-0}\cline{3-3}

{\begin{align*}
1.&&h(z,u)&\eqs h(u,z)\\
2.&&f(g(y),y)&\eqs g(h(x,y))\\
\end{align*}}
&

{\begin{align*}
\\
...\\
\end{align*}}

&

{\begin{align*}
1.&&h(z,u)&\eqs h(u,z)\\
2.&&f(g(y),y)&\eqs g(h(x,y))\con{x\succ y}\\
3.&&f(g(y),y)&\eqs g(h(y,x))
\end{align*}}

\\\cline{0-0}\cline{3-3}

\end{tabular}
\end{center}
We first obtain $S_1$ by performing the following superposition:
\begin{prooftree}
\AxiomC{$h(z,u)\eqs h(u,z)$}
\AxiomC{$f(g(y),y)\eqs g(h(x,y))$}
\BinaryInfC{$f(g(y),y)\eqs g(h(y,x))$}
\end{prooftree}
Then, we perform a demodulation step with the following redundancy formula:
$$\exists z,u.\,h(z,u)=h(x,y)\land h(z,u)\succ h(u,z)\land (f(g(y),y)\eqs g(h(x,y)))\succ h(z,u)$$
This simplifies to $x\succ y$, which we add to clause 2 in $S_1$ to obtain $S_2$. The two sides of the equation in clause 2 are incomparable, so it is possible to perform superpositions into both sides and with both orientations of the equation in $S_0$ and $S_1$.

However, for any substitution $\sigma$ s.t. $\sigma\vDash g(h(x,y))\succ f(g(y),y)$, we also have $\sigma\vDash x\succ y$. Hence, when using clause 2 oriented right-to-left in a \Sup{} inference in $S_2$,
\begin{enumerate}
\item if clause 2 is the left premise, then either condition (3) or (7) of \Sup{} is violated, and
\item if clause 2 is the right premise, then either condition (4) or (8) of \Sup{} is violated.
\end{enumerate}
For example, we skip the following superposition with clause 2 into itself in $S_2$:
\begin{prooftree}
\AxiomC{$f(g(y),y)\eqs g(h(x,y))$}
\AxiomC{$f(g(y),y)\eqs g(h(x,y))$}
\BinaryInfC{$f(f(g(y),y),h(x,y))\eqs g(h(z,h(x,y)))$}
\end{prooftree}
\vspace{-1.5em}\QED
\end{example}
For the next example, we use the following lemma.
\begin{restatable}{Lemma}{joinabilityLemma}\label{lem:trivial-joinability}%
Let $S_0, \ldots, S_i$ be a saturation s.t. there is a clause $C\con{R}$ in $S_i$ and for each substitution $\theta$ grounding for $C\con{R}$, either $\sigma\vDash R$, or $\sigma\vDash s=t$ for some equation $s\eqs t$ in $C$. Then $\top$ is admissible for $S_i$ and $C\con R$.
\end{restatable}
\fullversion{
\begin{proof}
If all substitutions $\theta$ grounding $C$ are such that $\theta \models R$ then the result follows by~\Cref{lem:R:means:redundant}. Otherwise, for all substitutions $\theta$ grounding $C$ such that $\theta \models R$ the reasoning is the same, so take any substitution $\theta$ grounding $C$ s.t. $\theta\nvDash R$. There is some equation $s\eqs t$ in $C$ s.t. $\theta\vDash s=t$. Then, $C\theta$ is redundant w.r.t. any set of clauses. Hence, $\top$ is admissible for $S_i$ and $C\con R$.\QED
\end{proof}
}
\noindent
The next example illustrates ground joinability~\cite{Duarte2022} using \Calc{}.
\begin{example}
\label{ex:ground-joinability}
Consider the following saturation:
\begin{center}
\small
\def\arraystretch{1.2}
\setlength{\abovedisplayskip}{0pt}
\setlength{\belowdisplayskip}{0pt}
\setlength{\abovedisplayshortskip}{0pt}
\setlength{\belowdisplayshortskip}{0pt}
\begin{tabular}{|>{\centering\arraybackslash}p{.4\linewidth}|>{\centering\arraybackslash}p{.03\linewidth}|>{\centering\arraybackslash}p{.5\linewidth}|}

\cline{0-0}\cline{3-3}
{\normalsize $S_0$} & & {\normalsize $S_2$}
\\\cline{0-0}\cline{3-3}

{\begin{align*}
1.&&f(z,u)&\eqs f(u,z)\\
2.&&g(f(x,a))&\eqs g(f(a,x))\\
\end{align*}}

&

{\begin{align*}
\\
...\\
\end{align*}}

&

{\begin{align*}
1.&&f(z,u)&\eqs f(u,z)\\
2.&&g(f(x,a))&\eqs g(f(a,x))\con x\succ a\\
3.&&g(f(a,x))&\eqs g(f(a,x))
\end{align*}}

\\\cline{0-0}\cline{3-3}

\end{tabular}
\end{center}
We perform the following superposition to obtain $S_1$:
\begin{prooftree}
\AxiomC{$f(z,u)\eqs f(u,z)$}
\AxiomC{$g(f(x,a))\eqs g(f(a,x))$}
\BinaryInfC{$g(f(a,x))\eqs g(f(a,x))$}
\end{prooftree}
$S_2$ is then obtained by adding $x\succ a$ to clause~2 using~\Cref{lem:demodulation}. We continue with the following clause sets:
\begin{center}
\small
\def\arraystretch{1.2}
\setlength{\abovedisplayskip}{0pt}
\setlength{\belowdisplayskip}{0pt}
\setlength{\abovedisplayshortskip}{0pt}
\setlength{\belowdisplayshortskip}{0pt}
\begin{tabular}{|>{\centering\arraybackslash}p{.42\linewidth}|>{\centering\arraybackslash}p{.03\linewidth}|>{\centering\arraybackslash}p{.5\linewidth}|}

\cline{0-0}\cline{3-3}
{\normalsize $S_3$} & & {\normalsize $S_5$}
\\\cline{0-0}\cline{3-3}

{\smaller
\begin{align*}
1.&&f(z,u)&\eqs f(u,z)\\
2.&&g(f(x,a))&\eqs g(f(a,x))\con x\succ a\\
3.&&g(f(a,x))&\eqs g(f(a,x))\con \top\\
\end{align*}}

&

{\begin{align*}
\\
...\\
\end{align*}}

&

{\smaller
\begin{align*}
1.&&f(z,u)&\eqs f(u,z)\\
2.&&g(f(x,a))&\eqs g(f(a,x))\con x\succ a\lor a\succ x\\
3.&&g(f(a,x))&\eqs g(f(a,x))\con \top\\
4.&&g(f(x,a))&\eqs g(f(x,a))
\end{align*}}

\\\cline{0-0}\cline{3-3}

\end{tabular}
\end{center}
For clause 3 in $S_2$, we have $\sigma\vDash g(f(a,x))=g(f(a,x))$ for any substitution $\sigma$, so we change the redundancy formula of clause 3 to $\top$ by~\Cref{lem:trivial-joinability} to obtain $S_3$. We perform a similar superposition as the first and attach corresponding constraints to obtain $S_5$. Finally, we change the redundancy formulas of both clauses 2 and 4 to $\top$, again by~\Cref{lem:trivial-joinability}. Hence, we obtain a clause set where all clauses except for clause 1 have redundancy formulas $\top$ and are identified as redundant.\QED
\end{example}






\section{Implementation and Evaluation}\label{sec:vampire}

\paragraph{Implementation.} We implemented a variant of saturation in the \vampire{} prover~\cite{CAV13} to demonstrate the applicability of the \Calc{} calculus. Our implementation uses a standard saturation algorithm, and partial clauses as follows:
\begin{enumerate}[itemsep=2pt,leftmargin=1.4em]
\item At the beginning of saturation, every clause $C$ is initialized as $C\con\bot$.
\item We perform \Calc{} inferences instead of standard superposition inferences, and as usual, skip inferences whose side conditions are violated.
\item If a \Calc{} inference is performed, we apply a redundancy step similar to demodulation steps (\Cref{sec:examples}) and add a redundancy formula to the main premise of the inference.
\end{enumerate}
We note that the added redundancy formulas might be slightly different from the one used in demodulation steps, corresponding to various demodulation variants used in \vampire{}, controlled by options.

Fully checking \Calc{} side conditions from Figure~\ref{fig:calculus} and handling redundancy formulas can be expensive. Our implementation uses several sufficient conditions to check these. Namely, we always checks side conditions (1)--(6) of \Sup{} inferences, side condition (1) of {\sf EqRes} inferences and side conditions (1)--(4) of {\sf EqFac} inferences. Additionally, when adding redundancy formulas to partial clauses, we simplify these redundancy formulas, similarly as in the examples of~\Cref{sec:examples}.

We  introduce the following Boolean options in \vampire{} to control \Calc:
\begin{enumerate}[itemsep=2pt,leftmargin=1.4em]
    \item The option {\tt -crc} toggles checking \emph{all} side conditions of the \Calc{} inferences, as well as attaching redundancy formulas to clauses when they contain tautological ordering constraints.
    \item The option {\tt -croc} depends on {\tt -crc} being enabled, and allows adding redundancy formulas with non-tautological ordering constraints.
    \item The option {\tt -crs} depends on {\tt -croc} being enabled, and toggles a more expensive case analysis on non-tautological ordering constraints, as in~\Cref{ex:redundant-orientation,ex:ground-joinability}.
\end{enumerate}
Our implementation  in \vampire{} requires less than a thousand lines of code\footnote{\url{https://github.com/vprover/vampire/tree/7deb7a5e2f08674992b6}}, as it utilizes existing efficient data structures: \emph{code trees}~\cite{PartiallyAdaptiveCodeTrees} for checking equality constraints and \emph{term ordering diagrams} for checking ordering constraints.\footnote{To be published.}
\paragraph{Experimental Setup.}
Our experiments used benchmarks from version 8.2.0 of the TPTP library~\cite{TPTP}, and were run using AMD Epyc 7502 2.5GHz processors and 1TB RAM. Each benchmark ran with a single core and 16GB of memory.

For our evaluation, we used a Spider-style strategy discovery~\cite{SpiderStyleStrategyDiscovery}. This approach takes a set of problems (the \emph{training set}), discovers sets of options (\emph{strategies}) that solve these problems, and   finds a sequence of strategies (\emph{schedule}) that solves the most of these problems given an instruction limit. 
Our training set consists of 5000 randomly selected CNF problems with equality from TPTP. The strategy discovery phase lasted 7 days, and the strategies collected solved 3300 of the 5000 problems. The collected strategies were then used to create two schedules\footnote{\url{https://gist.github.com/mezpusz/cf07dbda3ee0c5eabebf391254222a70}} that greedily cover as many problems as possible:
\begin{enumerate}[itemsep=2pt,leftmargin=1.4em]
    \item the {\tt baseline} schedule contains 285 strategies, all of them with {\tt -crc}, {\tt -croc} and {\tt -crs} disabled,
    \item the {\tt modified} schedule has 342 strategies, containing 100 strategies with {\tt -crc} enabled, 29 strategies with {\tt -croc}, and 13 strategies with {\tt -crs}.
\end{enumerate}
We selected a \emph{test set} to test the {\tt baseline} and {\tt modified} schedules. This set consists of selected TPTP benchmarks that are not in the training set and do not yield immediate errors (for example, higher-order benchmarks). We ran both schedules on the training set and the test set, with 300 seconds timeout. Note that neither of the two schedules are fully executed within 300 seconds.

\begin{table}[t]
\caption{Results for running the {\tt baseline} and {\tt modified} schedules for 300 seconds on the training and test sets.}
\smaller
\centering
\renewcommand*\arraystretch{1.1}
\setlength{\tabcolsep}{0.4em}
\begin{tabular}{c|c|c|c|c|c|c|c|}
    \multirow{3}{*}{set} & \multirow{3}{*}{\begin{tabular}{c}number of\\benchmarks\end{tabular}} & \multicolumn{3}{c|}{{\tt baseline}} & \multicolumn{3}{c|}{{\tt modified}} \\
    \cline{3-8}
    & & \multirow{2}{*}{solved} & \multirow{2}{*}{uniques} & \multirow{2}{*}{\begin{tabular}{c}avg.rating\\of uniques\end{tabular}} & \multirow{2}{*}{solved} & \multirow{2}{*}{uniques} & \multirow{2}{*}{\begin{tabular}{c}avg.rating\\of uniques\end{tabular}} \\
    & & & & & & & \\
    \hline
    training & 5000 & 3216 & 22 & 0.82 & 3269 & 75 & 0.88 \\
    test & 15530 & 11612 & 94 & 0.61 & 11624 & 106 & 0.70 \\
    total & 20530 & 14828 & 116 & 0.65 & 14893 & 181 & 0.77
\end{tabular}
\label{tab:results}
\end{table}

Our results are summarized in \Cref{tab:results}. The {\tt modified} schedule solved 65 more problems, 53 in the training set and 12 in the test set. There were overall 181 uniques solved by the {\tt modified} schedule. We show the 16 highest ranking uniques solved in the two sets in~\Cref{tab:highest-rated-uniques}.
\begin{table}[t]
\caption{16 highest-rated problems in the training and test sets, respectively, solved by the {\tt modified} schedule and not solved by the {\tt baseline} schedule. Problems with {\tt Unknown} status are marked with $\dag$, the rest of the problems have status {\tt Theorem}/{\tt Unsatisfiable}.}
\smaller
\centering
\begin{minipage}{.48\linewidth}
\renewcommand*\arraystretch{1.1}
\setlength{\tabcolsep}{0.5em}
\begin{tabular}{l|c|l|c|}
\multicolumn{4}{c|}{Training set}\\
\hline
Problem & Rating & Problem & Rating\\
\hline
{\tt CAT030+3}$^\dag$ & 1.0 & {\tt REL040+3} & 1.0\\
{\tt KLE130+1} & 1.0 & {\tt REL040+4} & 1.0\\
{\tt KLE156+2} & 1.0 & {\tt REL040-4} & 1.0\\
{\tt LCL148-1} & 1.0 & {\tt SET368-6} & 1.0\\
{\tt NUM090-1} & 1.0 & {\tt SET546-6} & 1.0\\
{\tt NUM636+4} & 1.0 & {\tt SEU443+2} & 1.0\\
{\tt NUM660+4} & 1.0 & {\tt SWV276-1} & 1.0\\
{\tt NUM786+4} & 1.0 & {\tt SWV978-1} & 1.0
\end{tabular}
\end{minipage}\begin{minipage}{.52\linewidth}
\renewcommand*\arraystretch{1.1}
\setlength{\tabcolsep}{0.5em}
\begin{tabular}{|l|c|l|c}
\multicolumn{4}{|c}{Test set}\\
\hline
Problem & Rating & Problem & Rating\\
\hline
{\tt HAL001+1} & 0.94 & {\tt ALG232+4} & 1.0\\
{\tt LCL275-3} & 0.94 & {\tt CAT030+2}$^\dag$ & 1.0\\
{\tt NUN075+1} & 0.94 & {\tt HWV088\_1} & 1.0\\
{\tt CSR036-10} & 0.95 & {\tt ITP265\_3} & 1.0\\
{\tt LCL775-1} & 0.95 & {\tt ITP285\_4} & 1.0\\
{\tt REL040-3} & 0.95 & {\tt ITP382\_1} & 1.0\\
{\tt SET039-3} & 0.95 & {\tt LAT368+1} & 1.0\\
{\tt LAT357+2} & 0.97 & {\tt LCL646+1.015} & 1.0
\end{tabular}
\end{minipage}
\label{tab:highest-rated-uniques}
\end{table}

We show more detailed statistics regarding \Calc{} inferences of {\tt modified} runs in \Cref{tab:skipped-superpositions}. The table displays for each option the distribution of ratios $n/m$ for successful strategy runs with at least one \Calc{} inference performed, where $n$ is the number of discarded \Calc{} inferences due to the corresponding option being enabled, and $m$ is the number of performed \Calc{} inferences within that run. For example, there were 1313 successful strategy runs where {\tt -crc} was enabled and at least one \Calc{} inference was performed, 514 of them did not discard any \Calc{} inferences, while 21 discarded more than 1.6 times more \Calc{} inferences than they performed.

Below, we list some benchmarks that {\tt baseline} failed to solved, but {\tt modified} succeeded by using our new techniques:
\begin{itemize}[itemsep=2pt,leftmargin=1.2em]
\item {\tt LAT243-1} (rating 0.73) was solved by performing only 22,533 inferences and discarding 238,768 inferences due to {\tt -crc},
\item {\tt GRP177-1} (rating 0.95) was solved by performing 59,243 inferences and discarding 45,839 and 39,516 inferences due to {\tt -crc} and {\tt -croc}, respectively,
\item {\tt SWV021-10} (rating 0.67) was solved by performing only 11 inferences and discarding 18, 2 and 17 inferences due to {\tt -crc}, {\tt -croc} and {\tt -crs}, respectively.
\end{itemize}
\begin{table}[t]
    \caption{Distribution of ratios $n/m$ for successful strategy runs with at least one \Calc{} inference performed, where $n$ is the number discarded \Calc{} inferences due to the corresponding option being enabled ({\tt -crc}, {\tt -croc} or {\tt -crs}), and $m$ is the number of performed \Calc{} inferences within that run.}
    \centering
    \renewcommand*\arraystretch{1.1}
    \setlength{\tabcolsep}{0.5em}
    \begin{tabular}{c|c|c|c|c|c|c|c|c}
        option & total & $0$ & (0,0.1] & (0.1,0.2] & (0.2,0.4] & (0.4,0.8] & (0.8,1.6] & $>1.6$\\
        \hline
        {\tt -crc} & 1313 & 514 & 433 & 109 & 97 & 95 & 44 & 21\\
        {\tt -croc} & 194 & 119 & 47 & 8 & 20 & 4 & 0 & 0\\
        {\tt -crs} & 44 & 25 & 11 & 2 & 1 & 3 & 2 & 0
    \end{tabular}
    \label{tab:skipped-superpositions}
\end{table}
To summarize, our evaluation shows that our initial implementation of the \Calc{} calculus is complementary to standard superposition reasoning, and it solves many problems that could not be solved without \Calc{} reasoning.

\section{Related work}
\label{sec:related-work}

{Model construction}~\cite{BachmairGanzingerModelConstruction} is utilized in the majority of refutational completeness proofs for superposition-based calculi. Other notable approaches include {forcing}~\cite{PaisForcingCompleteness}, {semantic trees}~\cite{KirchnerDeductionWithSymbolicConstraints}, and using {closures} for {basic superposition}~\cite{BasicParamodulation} and {closure redundancy}~\cite{Duarte2022}. Model construction has been used for proving completeness of {higher-order superposition}~\cite{SuperpositionForFullHigherOrderLogic}, and it has several formalizations~\cite{SuperpositionFormalization,SuperCalc-AFP}. Standard redundancy and static and dynamic completeness have been  in detail discussed in~\cite{DBLP:books/el/RV01/BachmairG01,WaldmannSaturationFramework}.
\review{Our work proves static and dynamic completeness of \Calc{} at the same time, using a slightly non-standard notion of fairness and a derivation process that is based on redundancy formulas. Using redundancy formulas circumvents the need to separate redundancy for clauses and inferences, simplifying some reasoning in the proof.}

Constraints in first-order reasoning are mainly used to characterize subsets of ground instances of non-ground clauses~\cite{ConstraintsAndTheoremProving}. Unification constraints serve various purposes, from restricting the search space by implementing basicness restrictions~\cite{BasicParamodulation,BasicSuperpositionIsComplete,RedundancyCriteriaForConstrainedCompletion,KirchnerDeductionWithSymbolicConstraints} or dealing with infinitary unifiers~\cite{DelayedUnification,SuperpositionForFullHigherOrderLogic}, to separating first-order and theory reasoning~\cite{UWA}. Ordering constraints also enforce and strengthen ordering rewriting~\cite{TheoremProvingWithOrderingAndEqualityConstrainedClauses} and are implicitly used in ground joinability and  reducibility~\cite{Duarte2022,OnUSingGroundJoinableEquations,ground-reducibility}.
\review{Our redundancy formulas in \Calc{} contain predicates similar to unification and ordering constraints. \Calc{} differs from other constrained systems in that redundancy formulas are not inherited and they are compatible with standard redundancy elimination. This is due to the semantics of redundancy formulas, as they correspond to clauses on the ground level that are known to be redundant.}
Various algorithms and specialized data structures have been proposed to solve ordering constraints~\cite{SolvingSymbolicOrderingConstraints,KBOConstraints,LPOConstraintSolving,ConfluenceTrees,ThingsToKnowWhenImplementingKBO,EfficientCheckingOfTermOrderingConstraints}. Common axiomatizations, such as associativity-commutativity (AC), enjoy more specialized handling~\cite{OrderingsACSymbolicConstraints,OnUSingGroundJoinableEquations}.

More sophisticated restrictions on superposition inferences are special cases of connectedness~\cite{BachmairLeo1991Cep} and ground connectedness~\cite{Duarte2022}. These include, for example, {blocking}~\cite{CriticalPairCriteria}, {compositeness}~\cite{PrimeSuperposition} and {general superposition}~\cite{GeneralSuperposition,GeneralAndPrimeSuperpositions}.
\review{Our framework generalizes  such case analysis-based redundancy detection approaches since (partial) case analyses can be expressed as redundancy formulas. We  note that our partial redundancy method is not limited to unification and ordering predicates.}

\section{Conclusion}
We introduce partial redundancy to restrict quantified equational reasoning while retaining refutational completeness and full simplifications, resulting in the \Calc{} calculus. Our framework yields an alternative characterization of standard redundancy notions, using redundancy formulas, and opens up a new research direction in redundancy criteria. We implemented a variant of the \Calc{} calculus in the superposition theorem prover \vampire{}. Our evaluation demonstrates that partial redundancy is efficient and complements current techniques. Future work includes finding new ways to apply our framework within first-order reasoning and beyond, such as theories and higher-order logic. Developing tighter integration in saturation and designing more efficient data structures that fit our constraint solving purposes better is also a future work.

\subsubsection{Acknowledgements.}
We are grateful to Martin Suda for his help in strategy discovery and building the portfolio schedules.
This research was funded in whole or in part by the  ERC Consolidator Grant ARTIST 101002685, the ERC Proof of Concept Grant LEARN 101213411, the TU Wien Doctoral College SecInt, the FWF SpyCoDe Grant 10.55776/F85,  the WWTF grant ForSmart  10.47379/ICT22007, and the Amazon Research Award 2023 QuAT.


\bibliographystyle{splncs04}
\bibliography{bibliography}


\end{document}